\documentclass[reprint,amsmath,amssymb,aps,superscriptaddress,floatfix,pra]{revtex4-2}
\usepackage{siunitx}
\usepackage{graphicx}% Include figure files
\usepackage{dcolumn}% Align table columns on decimal point
\usepackage{bm}% bold math
\usepackage{soul}
\usepackage{color}
\usepackage{xcolor}
\colorlet{soulred}{red!30}
\sethlcolor{white}
\soulregister\cite7
\soulregister\ref7
\renewcommand\st[1]{}
\begin{document}

\preprint{AIP/123-QED}

\title{Creation of color centers in diamond by recoil implantation through dielectric films}% Force line breaks with \\
% \thanks{footnote}

\author{Yuyang Han}
\email{yh236@uw.edu}
\affiliation{Department of Electrical and Computer Engineering, University of Washington, Seattle, Washington 98195, USA}

\author{Christian Pederson}%
\affiliation{Department of Physics, University of Washington, Seattle, Washington 98195, USA}

\author{Bethany E. Matthews}%
\affiliation{Pacific Northwest National Laboratory, Richland, Washington 99352, USA}

\author{Nicholas S. Yama}%
\affiliation{Department of Electrical and Computer Engineering, University of Washington, Seattle, Washington 98195, USA}

\author{Maxwell F. Parsons}%
\affiliation{Department of Electrical and Computer Engineering, University of Washington, Seattle, Washington 98195, USA}

\author{Kai-Mei C. Fu}%
\affiliation{Department of Electrical and Computer Engineering, University of Washington, Seattle, Washington 98195, USA}
\affiliation{Department of Physics, University of Washington, Seattle, Washington 98195, USA}
\affiliation{Pacific Northwest National Laboratory, Richland, Washington 99352, USA}

\begin{abstract}
The need of near-surface color centers in diamond for quantum technologies motivates the controlled doping of specific extrinsic impurities into the crystal lattice. Recent experiments have shown that this can be achieved by momentum transfer from a surface precursor via ion \st{irradiation}\hl{implantation}, an approach known as ``recoil implantation.'' Here, we extend this technique to incorporate dielectric precursors for creating nitrogen-vacancy (NV) and silicon-vacancy (SiV) centers in diamond. Specifically, we demonstrate that gallium focused-ion-beam exposure to a thin layer of silicon nitride or silicon dioxide on the diamond surface results in the introduction of both extrinsic impurities and carbon vacancies. These defects subsequently give rise to near-surface NV and SiV centers with desirable properties after annealing.
\end{abstract}

%displayed desired
\maketitle

%\tableofcontents

In the past two decades, color centers in diamond have gained significant attention as atomic-scale sensors~\cite{mazeNanoscaleMagneticSensing2008,barryOpticalMagneticDetection2016,balasubramanianNanoscaleImagingMagnetometry2008} and as qubits for quantum information processing~\cite{fuExperimentalInvestigationQuantum2022,doldeHighfidelitySpinEntanglement2014a,bernienHeraldedEntanglementSolidstate2013a,delteilGenerationHeraldedEntanglement2016,hermansQubitTeleportationNonneighbouring2022}. The negatively charged nitrogen-vacancy (NV$^-$) center is particularly attractive for sensing due to its high internal quantum efficiency~\cite{mohtashamiQuantumEfficiencySingle2013}, room-temperature spin initialization~\cite{jacquesDynamicPolarizationSingle2009}, spin readout~\cite{robledoHighfidelityProjectiveReadout2011}, and long spin coherence time. The negatively charged silicon-vacancy (SiV$^-$) center shows promise for quantum network applications due to its stable, uniform, and strong optical transitions~\cite{rogersMultipleIntrinsicallyIdentical2014,neuSinglePhotonEmission2011} that are further protected from the electric field noise by its inversion symmetry~\cite{sipahigilIndistinguishablePhotonsSeparated2014}. However, the SiV$^-$ center also has drawbacks such as lower quantum efficiency and shorter spin coherence time~\cite{pingaultAllOpticalFormationCoherent2014}. Consequently, researchers are exploring alternative color centers in diamond such as other group IV defects, including germanium-vacancy (GeV), tin-vacancy (SnV), and lead-vacancy (PbV) centers~\cite{trusheimTransformLimitedPhotonsCoherent2020,siyushevOpticalMicrowaveControl2017, frochVersatileDirectwritingDopants2020}. To better study and engineer diamond defects for quantum technologies, means of controllably creating color centers in diamonds are of great interest. Particularly, the formation of very \st{near-suface}\hl{near-surface} defects (i.e.\,$<$\,10\,nm) is desirable for their potential applications in nanoscale sensing~\cite{rondinMagnetometryNitrogenvacancyDefects2014} and near-field photonics device coupling~\cite{janitzCavityQuantumElectrodynamics2020} for ultrafast single photon sources~\cite{hoangUltrafastSpontaneousEmission2015}.

The creation of these vacancy defect complexes (i.e., X-V centers) requires the presence of the extrinsic impurity X  -- either unintentionally or deliberately doped  -- and carbon vacancies, which can be introduced by electron, ion, or laser irradiation~\cite{ martinGenerationDetectionFluorescent1999a, koikeDisplacementThresholdEnergy1992,smithColourCentreGeneration2019}. Among the techniques to create X-V defects, X-ion implantation with post annealing has been widely used~\cite{meijerGenerationSingleColor2005, schroderScalableFocusedIon2017, pezzagnaCreationColourCentres2011, lagomarsinoCreationSiliconVacancyColor2021,CenterProductionSinglephoton}, as bombarding the diamond lattice with selected ions introduces both carbon vacancies and extrinsic impurities. Subsequent high-temperature anneal (usually above \SI{800}{\celsius}) allows vacancies to diffuse to the impurities and form X-V complexes~\cite{dhaenens-johanssonOpticalPropertiesNeutral2011}. 

During implantation, the lateral position of the ions can be controlled either by customized implantation masks or focused ion beam (FIB) direct writing~\cite{orwaUpperLimitLateral2012, schroderScalableFocusedIon2017,chakravarthiImpactSurfaceLaserinduced2021,toyliChipScaleNanofabricationSingle2010,baynGenerationEnsemblesIndividually2015}; the ion depth distribution by the implantation energy; and the ion density by the implantation dose, screening layers, and ion detection systems~\cite{meijerGenerationSingleColor2005,lagomarsinoCreationSiliconVacancyColor2021,itoNitrogenvacancyCentersCreated2017,pachecoIonImplantationDeterministic2017}. Near-surface NV centers, for example, that are 1\,-\,20\,nm below the diamond surface can be created by low energy (0.4\,-\,3\,keV) nitrogen ion beam implantation~\cite{ofori-okaiSpinPropertiesVery2012,sangtawesinOriginsDiamondSurface2019, maminNanoscaleNuclearMagnetic2013, staudacherNuclearMagneticResonance2013, phamNMRTechniqueDetermining2016} and nitrogen delta-doping~\cite{ohnoEngineeringShallowSpins2012}. Nevertheless, challenges remain in the creation of defects with sub-30\,nm accuracy and close-to-unity formation yield~\cite{bradacQuantumNanophotonicsGroup2019}. FIB implantation faces additional limitation in its ion source availability.

% For FIB implantation, the limitation upon the spacial accuracy comes from the effect of ion straggling, ion channeling, and the size of the implantation ion-beam. Formation yields are usually limited by the available vacancies around the implanted ions~\cite{schroderScalableFocusedIon2017,naydenovEnhancedGenerationSingle2010}. 

Recently, an alternative technique known as ``recoil implantation''~\cite{frochVersatileDirectwritingDopants2020,galeRecoilImplantationUsing2021, fischerRecoilImplantationThick1978, gailliardRecoilImplantationIon1984, grotzschelRecoilImplantationThin1978, paprockiRecoilImplantationAluminium1985} was used to address the limitation of ion sources by harnessing the momentum transfer from the ion-beam to the surface precursors, including solid-state films (silicon, germanium, tin, and lead)~\cite{frochVersatileDirectwritingDopants2020} and nitrogen-based gases (N$_2$, NF$_3$, and NH$_3$)~\cite{galeRecoilImplantationUsing2021}. To form vacancy complexes, a Xe FIB was used to displace the atoms from the surface precursors into the diamond lattice, creating both impurities and vacancies that give rise to the corresponding color centers after annealing. 

In this work, we extended the recoil implantation method to the use of dielectric precursors and gallium FIB. Specifically, we implanted near-surface nitrogen and silicon atoms by employing gallium FIB \st{irradiation}\hl{implantation} through a thin layer of either silicon nitride or silicon dioxide. Following high-temperature anneal and subsequent film removal, the formed NV and SiV centers \st{displayeded}\hl{displayed} desirable properties. This demonstration offers a more accessible approach to generating NV centers by recoil implantation, compared to using gas-phase precursors. In general, the extension of ion-source options and precursor materials made this technique more attractive for defect engineering.

\begin{figure*}[pbth]
    \centering
    \includegraphics[scale=1]{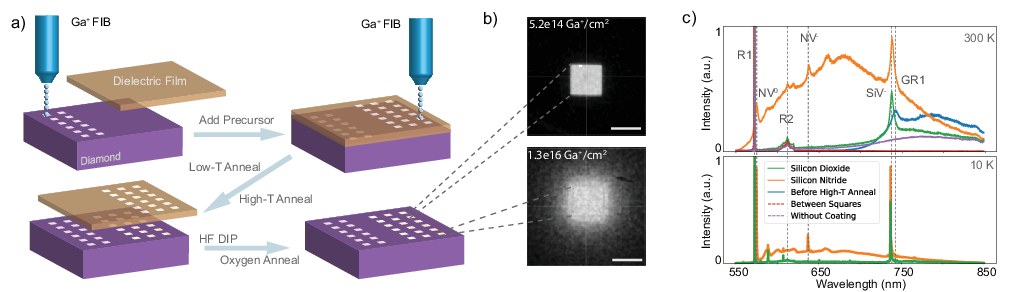}
    \caption{Experiment overview: \hl{(a)} illustration of the experiment process. The size of the \st{irradiation}\hl{implantation}-squares are not drawn to scale. The distance between the squares during each FIB exposure is $\SI{300}{\micro\metre}$. (b) Representative confocal images from two of the \st{irradiation}\hl{implantation} squares with different gallium fluence ($P_5$ for upper image and $P_1$ for lower image). Scale bars in the images are \st{$\SI{20}{\micro\metre}$}\hl{$\SI{10}{\micro\metre}$}. (c) Representative spectra at room temperature (upper plot) and at 10\,K (lower plot). The blue line represents data obtained from the highest-dose square of the SiO$_2$ sample prior to high-temperature annealing; all other spectra are acquired post high-temperature annealing. The red line is measured from a region between two squares; the purple line is taken from one of the squares irradiated with FIB prior to depositing the dielectric layers. Important wavelengths are marked by the grey dashed lines. In ascending order, these correspond to the first-order Raman line, NV$^0$ ZPL (575\,nm), second-order Raman line, NV$^-$ ZPL (637\,nm), SiV$^-$ ZPL (737\,nm), and GR1 ZPL (741\,nm).} 
    \label{outline}
\end{figure*}

Two $2\times2\times0.5$\,mm$^3$ single crystal type IIa chemical-vapor-deposition (CVD) diamond samples were used as substrates. The specified nitrogen impurity concentration is less than 5\,ppm (Element Six). Before the experiments, we cleaned both samples in acid and solvents (see Supplemental Material). Next, a control experiment was performed on both samples to introduce lattice damage in the absence of surface precursors. Each sample was exposed to 30\,keV Ga$^+$ ions in 12 $\SI{10}{\micro\metre}\times\SI{10}{\micro\metre}$ square regions. Six different FIB parameters were used for these 12 regions, with each parameter performed twice to test for variance between identical exposure conditions at different diamond surface locations (FIG.~\ref{outline}. (a)). The \st{irradiation}\hl{implantation} parameters, from $P_1$ to $P_6$, are summarized in TABLE \ref{doseTab}.

\begin{table}[htbp]
\begin{ruledtabular}
\begin{tabular}{llccl}
\textrm{Parameter}&
\textrm{I (nA)}&
\textrm{dwell ($\SI{}{\micro\second}$)}&
\textrm{Pass}&
\textrm{Fluence (cm$^{-2}$)}\\
\colrule
$P_{1}$ & 0.79  & 1      &  10   &   $(1.30\pm 0.14)\times 10^{16}$\\
$P_{2}$ & 0.79  & 1      &  2    &   $(2.61\pm 0.28)\times 10^{15}$\\
$P_{3}$ & 0.43  & 1      &  1    &   $(1.14\pm 0.15)\times 10^{15}$\\
$P_{4}$ & 0.08   & 1      &  1    &   $(9.7\pm 2.8)\times 10^{14}$\\
$P_{5}$ & 0.024  & 1      &  1    &   $(5.19\pm 0.50)\times 10^{14}$\\
$P_{6}$ & 0.0077 & 1      &  1    &   $(2.87\pm 0.36)\times 10^{14}$
\end{tabular}
\end{ruledtabular}
\caption{\label{doseTab} FIB parameters and estimated ion fluences: dwell time refers to how long the ion beam is at any location during one raster scan; pass represents the number of exposures performed on the same square region. 
}
\end{table}

The samples were then solvent-cleaned, after which we deposited a 5\,nm SiO$_2$ layer by electron-beam evaporation on one of the samples and a 5\,nm SiN$_x$ layer by plasma-enhanced CVD (PECVD) on the other. The deposited layers were used as surface precursors, where the same FIB exposure (TABLE~\ref{doseTab}) was performed to create 12 new \st{irradiation}\hl{implantation} squares on each sample (Fig.~\ref{outline}(a)). Six additional lower-dose squares were added to the SiO$_2$ sample for testing a wider range of ion fluences (see Supplemental Material.)

The samples were cleaned in acid after the second FIB exposure; the acid clean does not strip the dielectric films~\cite{williamsEtchRatesMicromachining2003} but removes contaminants prior to the \st{oxygen}\hl{low-temperature} anneal. This anneal, at $\SI{460}{\celsius}$ for four hours in an O$_2$ atmosphere, allows the carbon self-interstitials to diffuse~\cite{iakoubovskiiAnnealingVacanciesInterstitials2003}, thus healing some of the irradiation-induced damage \hl{(see Supplemental Material)}. However, the relatively low temperature does not lead to vacancy complex formation~\cite{greenDiamondSpectroscopyDefect2022}. Hence, this\st{ oxygen-anneal} is followed by a high-temperature anneal at $\SI{900}{\celsius}$ for 2 hours in a vacuum environment (see Supplemental Material). 25\% hydrofluoric acid (HF) was used to strip both dielectric layers after annealing. Finally, \hl{an}\st{a second} oxygen-anneal (see Supplemental Material) was performed on both samples to remove the irradiation-induced sp$^2$ carbon and oxygen terminate the surface for charge-state conversion to NV$^-$~\cite{fuConversionNeutralNitrogenvacancy2010}.

After the \st{first oxygen-anneal}\hl{low-temperature anneal} and before the high-temperature anneal, PL confocal imaging and spectroscopy of all \st{irradiation}\hl{implantation} squares was measured at room temperature with 532\,nm excitation. Representative confocal images are \st{displayeded}\hl{displayed} in FIG.~\ref{outline}(b), where a \hl{fluorescent square was found in each implanted region with the expected $\SI{10}{\micro\metre}\times\SI{10}{\micro\metre}$ size. An obvious}\st{visible} halo was observed outside each of the 3 highest-dose \st{irradiation} squares (\hl{with FIB parameters }$P_{1}$,$P_{2}$, and $P_{3}$)\st{,}\hl{. This was} attributed to neutralized gallium ions that were not fully focused. Spectra acquired from these fluorescent regions confirmed that the fluorescence comes from the GR1 center (neutral monovacancy defect) emission (FIG.~\ref{outline}(c)). Both the PL spectrum and intensity were consistent across the squares with identical FIB parameters, implying no intrinsic variation among different locations on the sample surface.

After the \st{second }oxygen-anneal, the PL emission changed from GR1 to NV and SiV$^-$. As shown in FIG.~\ref{outline}(c), the SiO$_2$ sample squares displayed a strong SiV$^-$ ZPL, while the SiN$_x$ sample squares had additional strong NV$^0$ and NV$^-$ ZPLs. This indicated that both the NV and SiV centers were created from the dielectric films. The control experiment confirmed that recoil implantation was the source of the doped nitrogen and silicon atoms, rather than diffusion from the dielectric films into the irradiated diamond during annealing. Additionally, a broadband irradiation-induced fluorescence was measured from all squares (FIG.~\ref{outline}(c), purple line). 

\begin{figure}[t]
    \centering
    \includegraphics[scale=1]{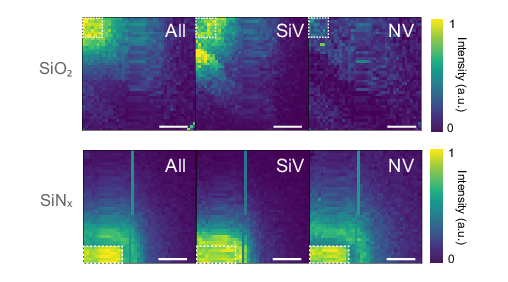}
    \caption{Hyperspectral images of the highest-dose \st{irradiation}\hl{implantation} squares from the SiO$_2$ sample and the SiN$_x$ sample. The excitation power was 4.2\,mW. From left to right, the intensity plots were obtained by: (All) integrating the total PL intensity over the 625\,nm-792\,nm confocal bandwidth; (SiV) binning the SiV$^-$ ZPL; and (NV) binning the NV$^-$ ZPL. Binning was performed by fitting and integrating over the corresponding ZPL (see Supplemental Material). The plots were not normalized to the same intensity. The \st{irradiation}\hl{implantation}-square boarder was marked by the white dashed line in each plot. The scale bars in the plots were $\SI{10}{\micro\meter}$.} 
    \label{hyperspec}
\end{figure}

Around the highest-fluence \st{irradiation}\hl{implantation} square from each sample, we took a hyperspectral image (FIG.~\ref{hyperspec}). Comparing the SiN$_x$ sample to the SiO$_2$ sample, the NV ZPL in the former was notably brighter within the irradiated region, while the latter had a more homogeneous intensity distribution. This confirmed that the NV centers in the SiN$_x$ sample arose from the nitrogen-based precursor, rather than from the naturally occurring impurities as in the SiO$_2$ sample.

\begin{figure}[htbp]
    \centering
    \includegraphics[scale=1]{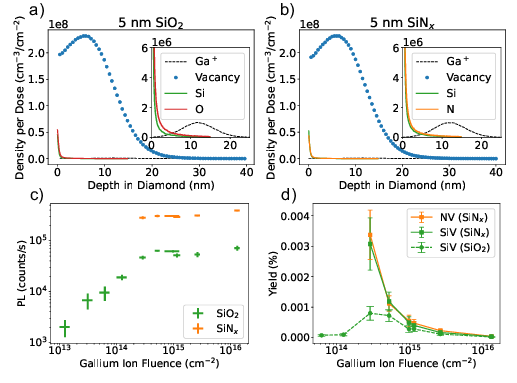}
    \caption{Formation yield estimate: TRIM simulations of 30\,keV incident Ga$^+$ beam on diamond with (a) 5\,nm electron-evaporation SiO$_2$ (2.658\,g/cm$^3$) and (b) 5\,nm PECVD SiN$_x$ (2.5\,g/cm$^3$). To derive the effective implantation dose of nitrogen and silicon atoms, integration over the simulation depth was performed, followed by multiplication with the gallium ion dose (Supplemental Material). This resulted in 1.83 silicon atom per gallium ion for the SiO$_2$ sample and 2.18 silicon plus 2.19 nitrogen atoms per gallium ion for the SiN$_x$ sample. (c) Confocal PL intensities for all \st{irradiation}\hl{implantation} squares measured at 0.2\,mW 532\,nm excitation. The vertical width of each data point showed the standard deviation of the measured PL intensities inside the square; the horizontal width showed the uncertainty in ion fluence. \hl{(d)} Formation yield for SiV and NV centers from the SiN$_x$ and SiO$_2$ samples that is computed based on (a), (b), and (c).} 
    \label{yield}
\end{figure}

\begin{figure*}[thbp]
    \centering
    \includegraphics[scale=1]{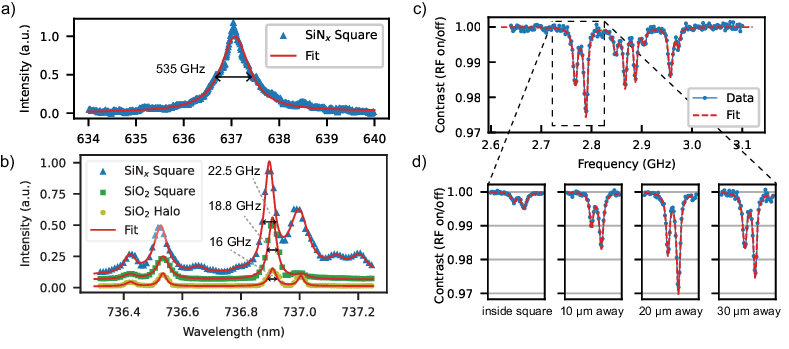}
    \caption{Defect characterization: (a) 10\,K PL spectrum of NV$^-$ centers from the SiN$_x$ sample. The spectrum was fit with a Lorentzian function to determine a FWHM of 535\,GHz. (b) 10\,K PL spectra of SiV$^-$ center(s) from both samples. The spectrum from an \st{irradiation}\hl{implantation} square on the SiN$_x$ sample (labeled ``SiN Square'') was fit by eight Lorentzian functions; the spectrum from a square on the SiO$_2$ sample (labeled ``SiO Square'') and the spectrum from the halo region around this square (labeled ``SiO Halo'') were each \st{Ffit}\hl{fit} with four Lorentzian functions. The spectra shown in (b) were scaled and shifted for better visualization. (c) representative ODMR spectrum from NV centers in the halo region of highest-fluence square on the SiN$_x$ sample. The same experiments were performed when the excitation beam is inside the square, $\SI{10}{\micro\meter}$, $\SI{20}{\micro\meter}$, and $\SI{30}{\micro\meter}$ away from the center of the square. The first two resonances from these measurements are shown in (d). The asymmetric spectrum intensity comes from the finite bandwidth of our microwave antenna (Supplemental Material).} 
    \label{characterize}
\end{figure*}

To calculate the formation yield of the NV and SiV centers, we first simulated the nitrogen and silicon atom distribution by TRIM~\cite{zieglerSRIMStoppingRange2010} and obtained the number of recoil-implanted atoms per gallium ion (FIG.~\ref{yield}(a) and (b)). Relative contributions to the total PL intensity (FIG.~\ref{yield}(c)) from the different color centers were then estimated by fitting the PL spectra with a weighted sum of NV$^0$ spectrum, NV$^-$ spectrum, SiV$^-$ spectrum, and the ``irradiation-induced fluorescence'' spectrum (FIG.~\ref{outline}(d), purple line). After normalizing the respective PL intensity to that from a sample with known defect density, we derived the formation yield (FIG.~\ref{yield}(d)) for NV and SiV centers in both samples (see Supplemental Material). The maximum yield was on the order of \st{$10^{-3}\%$}\hl{0.001\%}, much lower than 0.5\%\,-\,50\% that was obtained by direct ion implantation~\cite{lagomarsinoCreationSiliconVacancyColor2021,schroderScalableFocusedIon2017,spinicelliEngineeredArraysNitrogenvacancy2011,naydenovEnhancedGenerationSingle2010,sangtawesinHighlyTunableFormation2014,itoNitrogenvacancyCentersCreated2017, meijerGenerationSingleColor2005}. However, this was expected as the implanted atoms were not only very close to the surface but also in regions of very high ion fluence.\st{Both factors had the effect of reducing the PL intensity~\cite{santoriVerticalDistributionNitrogenvacancy2009} and decreasing the formation yield~\cite{schroderScalableFocusedIon2017, ofori-okaiSpinPropertiesVery2012}.} \hl{Very near-surface color centers are less fluorescent~\cite{santoriVerticalDistributionNitrogenvacancy2009,ofori-okaiSpinPropertiesVery2012}, and excessive ion fluence can damage the crystal lattice thus leading to low formation yield~\cite{schroderScalableFocusedIon2017, ofori-okaiSpinPropertiesVery2012}. Because the lattice damage increases with the ion fluence, the ``saturated'' PL intensities for the six highest-dose squares (FIG.~\ref{yield}(c)) implies that we have used excessive dose for all SiN$_x$ squares. However, it was still possible to probe lower-dose conditions by taking measurements in the halo region, where the fluence of the neutralized gallium ions should be much lower.} \st{In particular}\hl{As shown by FIG.~\ref{yield}(d)}, the \st{calculated }yield of the SiV centers was higher in the SiN$_x$ sample\st{,}\hl{.} \hl{This}\st{which} is likely because the SiV$^{2-}$ (dark state) \st{could}\hl{can} capture holes that \st{were}\hl{are} photo-generated from the nearby NV$^0$ centers and \st{became}\hl{become} SiV$^-$ (bright state)~\cite{gardillProbingChargeDynamics2021}. \st{The PL intensities in FIG.~\ref{yield}(c) remained ``saturated'' for the six highest-dose squares. This implied that we used excessive ion fluences for all of the SiN$_x$ squares. However, it was still possible to probe lower-dose conditions by taking measurements in the halo region, where the fluence of the neutralized gallium ions should be much lower.}

To characterize the optical properties of the formed defects, we measured the PL spectra from all \st{irradiation}\hl{implantation} squares at 10\,K. Most SiV centers from the squares exhibited four distinct main transition peaks, whereas the NV centers showed large inhomogeneous broadening (FIG.~\ref{characterize}(a) and (b)). Among the SiV$^-$ centers, we found the smallest inhomogeneous broadening (FWHM) of 16\,GHz from the halo region of the highest-dose square in the SiO$_2$ sample. This inhomogeneous linewidth was close to the spectrometer limit (10\,GHz); similar linewidths from several other \st{irradiation}\hl{implantation} squares were all comparable to the 15\,GHz linewidth observed in the literature by direct silicon ion implantation~\cite{evansNarrowLinewidthHomogeneousOptical2016}. Nevertheless, the recoil-implanted SiV centers were 10 times closer to the diamond surface (less than 10\,nm compared to around 100\,nm), which typically would lead to spectral diffusion due to the unfavorable surface strain and charge environment. \hl{In particular, other implanted defects such as the substitutional gallium, which is an acceptor~\cite{gossVacancyimpurityComplexesLimitations2005a}, and the nitrogen defect, which is a donor~\cite{santoriVerticalDistributionNitrogenvacancy2009}, will affect the equilibrium and dynamic charge state of the target impurity. Charging and discharging of these near-by defects can also contribute further to fluctuations in the electrostatic environment, although in some cases higher nitrogen can reduce such adverse fluctuations~\cite{orphal-kobinOpticallyCoherentNitrogenVacancy2023}.} 

We further investigated the spin coherence of the NV centers from the highest-dose square on the SiN$_x$ sample, by conducting continuous wave optically detected magnetic resonance (CW-ODMR). In the presence of a 24\,G external magnetic field, we observed four pairs of resonance dips that corresponded to the 4 different crystallographic orientations of NV centers (FIG.~\ref{characterize}(c)). As the excitation beam moved from inside this square to the halo region, the maximum PL contrast for a single resonance dip increased from 0.47\% to 2.94\%, while the average linewidth (FWHM) decreased from 10.3\,MHz to 8.37\,MHz (FIG.~\ref{characterize}(d)). This suggested that the spin-coherence of the NV centers would benefit from lower ion fluences than what we used in this study. Additionally, these linewidths were broadened by the 2\,mW excitation power~\cite{dreauAvoidingPowerBroadening2011}. We obtained the narrowest spectrum with a linewidth of 6.81\,MHz using 0.1\,mW excitation power (Supplemental Material). This observed 2.94\% contrast was approximately half of the expected maximum. This was due to the lower NV$^-$/NV$^0$ ratio and the additional background PL from the SiV$^-$ centers (Supplemental Material). The reduction in ODMR contrast served as a trade-off for having NV sensors at these shallow depths and might be improved by additional surface preparations\hl{.}

\begin{figure*}[pthb]
    \centering
    \includegraphics[scale=1]{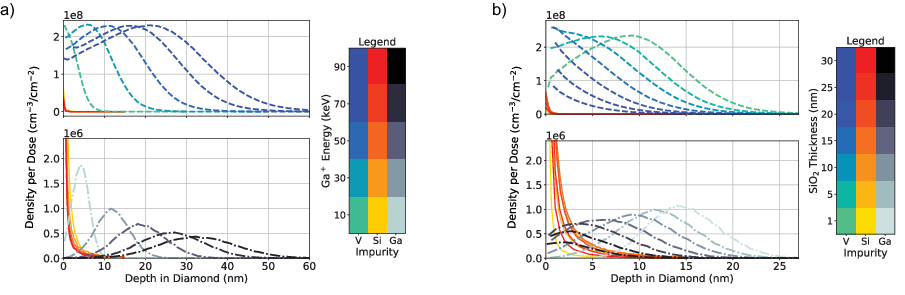}
    \caption{\st{TRIM simulations of Ga$^+$ ion irradiation on diamond with a SiO$_2$ layer on surface.}\hl{TRIM simulation of recoil implantation with variable ion energy and SiO$_2$ thickness.} (a) \hl{Impurity distribution using a}\st{A} 5\,nm SiO$_2$ layer with variable ion energies. \st{From small to large, the energies were 10\,keV, 30\,keV, 50\,keV, 70\,keV, and 90\,keV. Plot (c) was the zoom-in of plot (a).} (b) \hl{Impurity distribution using}\st{Using} 30\,keV gallium ion energy with variable SiO$_2$ layer thickness. \st{From small to large the thicknesses were 1\,nm, 5\,nm, 10\,nm, 15\,nm, 20\,nm, 25\,nm, and 30\,nm. Plot (d) was the zoom-in of plot (b)}. \st{The vertical z-axes of these plots represented density per dose (cm$^{-3}/$cm$^{-2}$).} \hl{The horizontal axis labeled ``Depth in Diamond (nm)" represents the depth measured from the diamond surface. The colors of the impurity distributions are represented in the legend of each plot.}} 
    \label{outlook}
\end{figure*}

Additional TRIM simulations were performed with an SiO$_2$ precursor to explore the flexibility provided by recoil implantation. As shown in Fig.~\ref{outlook}\st{(a) and (c)}, both the carbon vacancy and the silicon atom distributions \st{could}\hl{can} be adjusted by varying the ion energy and the SiO$_2$ layer thickness. This semi-independent control was not possible by direct silicon ion implantation, where the vacancy to silicon atom ratio was set by the implantation energy only. Potentially, this new degree of freedom allows for forming better defects by controlling the distribution of the gallium impurities. \st{In particular}\hl{For example, Fig.~\ref{outlook}(a) suggests that a large implantation energy sends the Ga$^+$ ions to a different depth from these silicon atoms. Alternatively,} Fig.~\ref{outlook}\st{(d)}\hl{(b)} \st{shows that using }\hl{implies that }a very thin layer of SiO$_2$ \hl{will achieve the same effect, while additionally realizing} \st{achieved} a much smaller range of depth for the silicon atoms.\st{, while sending the Ga$^+$ ions to a different depth from these silicon atoms.} This \st{also opened}\hl{opens} the possibility of forming a 2-d layer of defects on the crystal surface. In this case, the lateral straggle is expected to be small, which indicates that the spatial resolution is only limited by the ion beam-profile.

To conclude, we demonstrated gallium FIB recoil implantation through dielectric precursors for generating color centers in diamond. The use of silicon-dioxide and silicon-nitride films in this experiment expanded the range of applicable surface precursors to include dielectric materials, exemplified by the formation of near-surface SiV and NV centers. These SiV centers exhibited narrow inhomogeneous broadening that were promising for performing photoluminescence excitation and photonics device coupling, while the NV centers displayed reasonable ODMR spectra, highlighting their efficacy in quantum sensing applications. Simulations also underscored the flexibility of this technique, especially the degrees of freedom to fine-tune\st{ the} the spatial distributions and density profiles of the implanted atoms.

% \hl{ }

This material is based upon work \hl{primarily }supported by Department of Energy, Office of Science, National Quantum Information Science Research Centers, Co-design Center for Quantum Advantage (C2QA), under contract number DE-SC0012704\hl{, partially supported by NSF PHY-GRS 2233120}. The FIB implantation was performed at the Pacific Northwest National Laboratory. The dielectric film deposition was performed at the Washington Nanofabrication Facility, a National Nanotechnology Coordinated Infrastructure (NNCI) site at the University of Washington with partial support from the National Science Foundation via awards NNCI-1542101 and NNCI-2025489. The optical and ODMR measurements were performed in the Quantum Technologies Training and Testbed (QT3) user facility, supported by the National Science Foundation under award OMA-1936100 and OMA-1936932. The authors would like to thank Adam Cox and Chukwuemeka Mordi for their contributions in \st{developing the hyperspectral imaging scripts}\hl{software development}.

%\nocite{*}
\bibliographystyle{unsrt}
\bibliography{ma_Book}% Produces the bibliography via BibTeX.

\end{document}

\title{Supplemental Material for “Creation of color centers in diamond by recoil implantation through dielectric films”}

\author{Yuyang Han}
\email{yh236@uw.edu}
\affiliation{Department of Electrical and Computer Engineering, University of Washington, Seattle, Washington 98195, USA}

\author{Christian Pederson}%
\affiliation{Department of Physics, University of Washington, Seattle, Washington 98195, USA}

\author{Bethany E. Matthews}%
\affiliation{Pacific Northwest National Laboratory, Richland, Washington 99352, USA}

\author{Nicholas S. Yama}%
\affiliation{Department of Electrical and Computer Engineering, University of Washington, Seattle, Washington 98195, USA}

\author{Maxwell F. Parsons}%
\affiliation{Department of Electrical and Computer Engineering, University of Washington, Seattle, Washington 98195, USA}

\author{Kai-Mei C. Fu}%
\affiliation{Department of Electrical and Computer Engineering, University of Washington, Seattle, Washington 98195, USA}
\affiliation{Department of Physics, University of Washington, Seattle, Washington 98195, USA}
\affiliation{Pacific Northwest National Laboratory, Richland, Washington 99352, USA}

% \date{\today}

\maketitle

\section{Experiment Details}\label{details}

\subsection{Sample cleaning}
Two different methods were used in this work to clean the samples, which were referred to as  solvent clean (or ``cleaned in solvents'') and acid clean (or ``cleaned in acid'') in the main article. The solvent clean aimed to remove the dust-like contaminants that were usually visible under a 10$\times$ magnification desktop-microscope. This process included first sonicating the sample in acetone for 15 minutes and in isopropyl alcohol (IPA) for another 15 minutes, then drying the sample with nitrogen gas. Before every acid clean, a solvent clean was performed to avoid unwanted masking during the process. The acid-cleaning process included immersing the sample in a solution of 20\,ml sulfuric acid and 1\,g potassium nitrate at 225\,°C for one hour, then rinsing the samples in deionized water and drying with nitrogen gas.

\subsection{Focused ion beam \st{irradiation}\hl{implantation}}
\begin{figure}[ht!]
    \centering
    \includegraphics[scale=1]{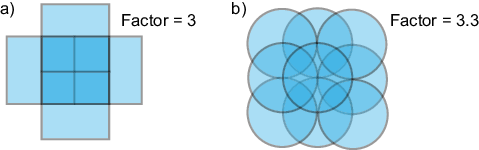}
    \caption{\hl{Illustration diagram for calculating the effective dose multiplication factor for a 50\% beam overlap. (a). By approximating the beam profile as a square at each step, then the factor of 3 comes from four 50\% dose contributions from each of the four neighbor steps (up, down, left, and right) plus a 100\% dose contribution by this center step per se. (b). Considering a more realistic case where the beam spot is approximated by a circle, the overall dose contribution can be calculated by the 8 spots around it (upper left, up, upper right, left, right, lower left, down, lower right) and the contribution from the center step per se. This gives a factor of $\frac{17}{3}-\frac{2 \left(\sqrt{3}+2\right)}{\pi }\approx3.3$.}} 
    \label{overlap}
\end{figure}

\begin{figure}[ht!]
    \centering
    \includegraphics[scale=1]{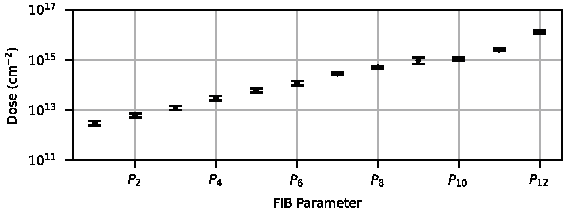}
    \caption{Calculated ion-dose and uncertainties for each FIB parameter.} 
    \label{dose}
\end{figure}

A Thermofisher Helios 660 Nanolab FIB/SEM Dual Beam system was used in this work to create the \st{irradiation}\hl{implantation} squares. The \st{irradiation}\hl{implantation} squares were created in the imaging mode by controlling the implantation energy, beam current, patterning overlap, and dwell time. Imaging of the sample surface was performed using scanning electron microscopy only to avoid unintentional FIB exposure. The patterning overlap referred to how much the ion beam profile overlapped with each other at each step during a raster scan. Throughout the FIB experiments, 30\,keV energy and 50\% overlap were used. The focused beam diameters for each beam current (I) were estimated by relations provided in Ref.~\cite{bassimRecentAdvancesFocused2014}. To calculate the effective ion dose, a factor of 3 was multiplied to account for the 50\% overlap\hl{ (see FIG.~\ref{overlap})}, and the number of passes was multiplied to consider the number of \st{irradiation}\hl{implantation} performed at the same location. FIB parameters used in this work were summarized in Table \ref{doseTab} and plotted (with uncertainty) in a log scale in FIG.~\ref{dose}. The Uncertainties came from the beam diameter estimate.

\begin{table}[ht!]
\caption{
FIB Parameters and Estimated Doses}
\label{doseTab}%
\begin{tabular}{llcccl}
\textrm{Parameter}&
\textrm{I (nA)}&
\textrm{dwell time ($\SI{}{\micro\second}$)}&
\textrm{Pass}&
\textrm{\hl{Beam Diameter (nm)}}&
\textrm{Dose (cm$^{-2}$)}\\
$P_{1}$ & 0.79  & 1      &  10  &  \hl{$38\pm 2$}  &   $(1.30\pm 0.14)\times 10^{16}$\\
$P_{2}$ & 0.79  & 1      &  2  &  \hl{$38\pm 2$}  &   $(2.61\pm 0.28)\times 10^{15}$\\
$P_{3}$ & 0.43  & 1      &  1  &  \hl{$30\pm 2$}  &   $(1.14\pm 0.15)\times 10^{15}$\\
$P_{4}$ & 0.08   & 1      &  1  &  \hl{$14\pm 2$}  &   $(9.7\pm 2.8)\times 10^{14}$\\
$P_{5}$ & 0.024  & 1      &  1  &  \hl{$10.5\pm 0.5$}  &   $(5.19\pm 0.50)\times 10^{14}$\\
$P_{6}$ & 0.0077 & 1      &  1  &  \hl{$8\pm 0.5$}  &   $(2.87\pm 0.36)\times 10^{14}$\\
$P_{7}$ & 0.0013 & 1      &  1  &  \hl{$5\pm 0.5$}  &   $(1.24\pm 0.25)\times 10^{14}$\\
$P_{8}$ & 0.0013 & 0.5    &  1  &  \hl{$5\pm 0.5$}  &   $(6.2\pm 1.2)\times 10^{13}$\\
$P_{9}$ & 0.0013 & 0.25   &  1  &  \hl{$5\pm 0.5$}  &   $(3.10\pm 0.62)\times 10^{13}$\\
$P_{10}$ & 0.0013 & 0.1    &  1  &  \hl{$5\pm 0.5$}  &   $(1.24\pm 0.25)\times 10^{13}$\\
$P_{11}$ & 0.0013 & 0.05   &  1  &  \hl{$5\pm 0.5$}  &   $(6.2\pm 1.2)\times 10^{12}$\\
$P_{12}$ & 0.0013 & 0.25   &  1  &  \hl{$5\pm 0.5$}  &   $(3.10\pm 0.62)\times 10^{12}$
\end{tabular}
\end{table}

\subsection{Sample annealing}

Two different annealing setups were used in this work, which were referred to as ``oxygen anneal\hl{ (low-temperature anneal)}'' and ``high-temperature anneal'' in the main article. 

\begin{figure}[ht!]
    \centering
    \includegraphics[scale=1.5]{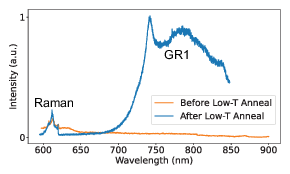}
    \caption{\hl{Spectral change before and after the low-temperature anneal inside the same implantation square: before the anneal, no GR1 emission was observed, and the second-order diamond Raman line was reduced compared to the PL spectrum outside the square (pure diamond). The implantation squares were visible under the confocal microscope due to a broadband fluorescence from 600\,nm to 900\,nm. After low-temperature annealing, the squares became more fluorescent. Spectrally, the GR1 spectrum (741\,nm zero-phonon line and the higher-wavelength phonon sideband) appeared; the Raman intensity was recovered; and the broadband fluorescence was removed.}} 
    \label{change}
\end{figure}

During an oxygen anneal\hl{ (low-temperature anneal)}, the samples were first heated up to 460\,°C in 30 minutes and stays at 460\,°C for two hours. The samples were exposed to an oxygen atmosphere during the entire process for selective oxidation which removed the irradiation-induced sp$^2$ carbons on the surface and helped to convert the charge state of NV$^0$ centers to NV$^-$~\cite{fuConversionNeutralNitrogenvacancy2010,rondinSurfaceinducedChargeState2010}. In this work, \st{oxygen}\hl{the low-temperature} anneal was also used to \st{partially heal the lattice damage since carbon self-interstitials start to diffuse at 400\,°C, which made}\hl{make} visible the strong GR1 line before the high-temperature anneal\hl{ (FIG.~\ref{change}). Because this anneal was conducted at a relatively low temperature (460 °C) where only the interstitial carbon atoms were mobile (above 400 °C), we expect this spectral change to be a result of partial healing the lattice damage. However, there was no direct evidence that confirmed this theory. An alternative explanation would be oxygen termination on the diamond surface, but this is unlikely because the surface was coated with the dielectric layer during the entire annealing process}. 

\begin{figure}[ht!]
    \centering
    \includegraphics[scale=1]{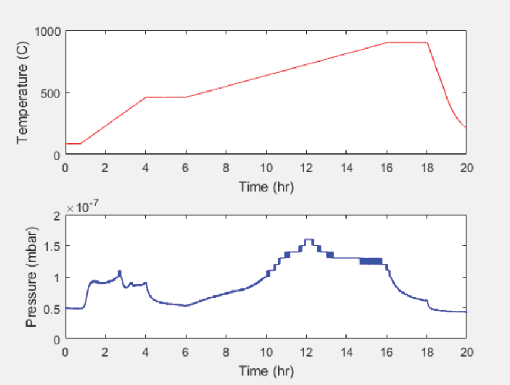}
    \caption{Temperature and pressure change during the high-temperature anneal.}
    \label{anneal}
\end{figure}

The high-temperature anneal was conducted in a vacuum environment, where the temperature and pressure were monitored during the process. As shown in Fig.~\ref{anneal}, the anneal started with a pressure of 5$\times$10$^{-8}$\,mbar, which raised to 1$\times$10$^{-7}$\,mbar as we heated up the sample to 460\,°C. Therefore, We programmed the temperature to stay at  460\,°C for two hours until a drop in the pressure was observed before rising to the 900\,°C annealing temperature over 10 hours. The samples were held at 900\,°C for two hours. This anneal provided enough energy for the vacancies to diffuse and form vacancy complexes~\cite{greenDiamondSpectroscopyDefect2022}.

\section{Hyperspectral Binning}\label{fitting}

\begin{figure}[ht!]
    \centering
    \includegraphics[scale=1]{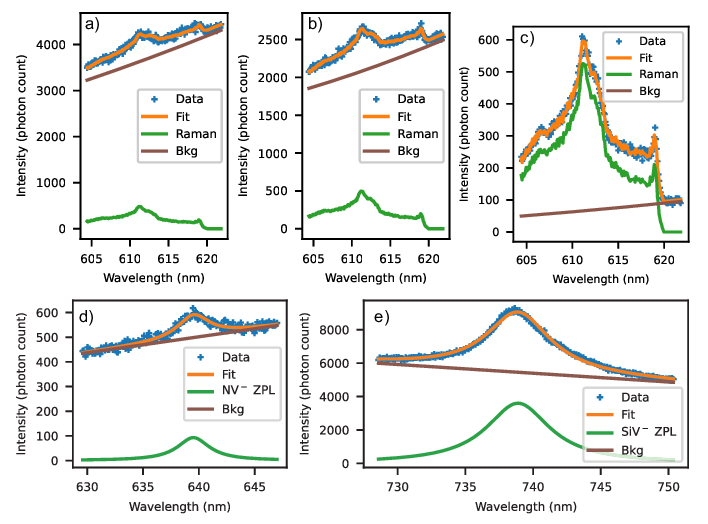}
    \caption{Representative fits to separate the second-order Raman line (a)\,-\,(c), NV$^-$ zero-phonon line (d), and SiV$^-$ zero-phonon line (e) from their background. (a), (b), and (c) showed three second-order Raman spectra affected by different amount of NV$^0$ PSB background. The unmodified diamond Raman spectrum profile was obtained from a non-irradiated region (main article FIG.~1(c), red line).} 
    \label{r2fit}
\end{figure}

Because there were more than one kind of defect created at the same location in this work, the PL spectrum comprised overlapping spectra of lower-wavelength defects' phonon sidebands (PSB) and the higher-wavelength defects' zero-phonon lines (ZPL). For this reason, binning the hyperspectral data was more complicated than simply integrating over the wavelength of interest. For instance, integrating over a 1\,nm window of spectrum intensity around the SiV$^-$ ZPL would automatically take the NV$^-$ PSB into consideration, which could dominate the result in some cases, as shown by the representative SiN$_x$ spectrum in FIG.~1(c) of the main article. However, in the vicinity of the ZPL peak, its background from other defects could be approximated by a linear or a quadratic function. Therefore, for binning NV$^-$ and SiV$^-$ fluorescence, we fitted the sum of a quadratic background and a Lorentzian peak to spectrally separate the corresponding ZPL from other contributors, followed by integrating over the Lorentzian component (FIG.~\ref{r2fit}(d) and (e)). This method worked as intended, which can be seen by binning the second-order Raman line (FIG.~\ref{r2fit}(a)\,-\,(c)), where the spectrum profile was no longer a single peak.

\section{Yield Calculation}\label{yield}

\subsection{PL contribution}
\begin{figure}[ht!]
    \centering
    \includegraphics[scale=1.2]{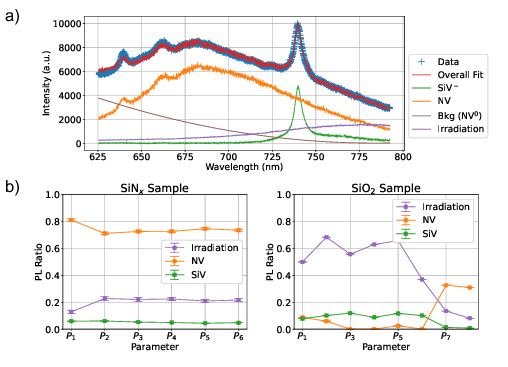}
    \caption{(a) Representative fits from a SiN$_x$ sample PL spectrum to separate the components of the NV emission, SiV$^-$ emission, an irradiation-induced emission, and a quadratic background. (b) The relative PL intensity ratio in each sample that was derived form the fits shown in (a).} 
    \label{ratio}
\end{figure}

In order to calculate the formation yield, it was necessary to separate the SiV and NV emissions from the PL of each irradiation square and estimate the number of defects inside the excitation beam. Similar to separating the second-order Raman line from its background, this was done by first obtaining the spectrum model for each component that consists of the total PL and fitting the overall spectrum (inside our detection bandwidth, 625\,nm-to-792\,nm) by scaling the intensity of each model spectrum. Model spectra and the data were obtained at room temperature with the same optical setup. The final model consisted of an NV spectrum, an SiV$^-$ spectrum, an irradiation-induced spectrum (main article, FIG.1(c), purple line), and a quadratic background. The quadratic background was used to compensate for the difference of the NV$^-$/NV$^0$ ratio (see Section~\ref{charge}) among different samples and different surface locations. The NV model spectrum was obtained from a N$^+$ implantation sample, and the SiV$^-$ model was obtained by subtracting the NV model from the SiN$_x$ sample spectrum. As shown in FIG.~\ref{ratio}, scaling the intensities of the models was sufficient to fit the total spectrum accurately. From there, the spectrum intensity for each component was integrated over the detection bandwidth to calculate the relative contributions to the total PL intensity from each defect. In particular, the NV PL intensity was calculated by adding the background spectrum to the NV model spectrum, since the background intensity was mostly (if not all) contributed by the PSB of NV$^0$ emission. Because the PL intensities were measured by the confocal setup that had a specific detection bandwidth, these results did not represent the absolute fluorescence intensity ratio of the defects (i.e. in all wavelengths). Further, because the SiV and NV fluorescence had different power-dependence, this calculation could only be applied to the specific excitation power (0.2\,mW) used in the measurements.

\subsection{Intensity normalization}
Additional calibration for each component's PL intensity was conducted based on the fitting results, in order to calculate the number of NV and SiV centers inside the excitation beam. The NV PL intensity was normalized with respect to the PL intensity of the NVs in bulk ($>$\,50 micron), where 20 single NVs' PL intensities were averaged to account for the intensity difference among different crystallographic orientations. The number of SiV centers was calibrated with respect to the PL intensity from the SiV center ensembles in a sample with 30\,keV 1e11 cm$^{-2}$ Si$^+$ ion implantation and 2-hour \SI{1200}{\celsius} post-annealing. For this sample, we assumed a SiV formation yield of 5\%~\cite{lagomarsinoCreationSiliconVacancyColor2021,schroderScalableFocusedIon2017,bradacQuantumNanophotonicsGroup2019} to estimate the average single SiV$^-$ PL intensity. 

Uncertainties were introduced at each step of the yield calculation due to the approximations made. Specifically, the estimated effective nitrogen/silicon implantation dose from the TRIM simulations did not exclude atoms that are within lattice constants from the diamond surface, where NV and SiV centers are not likely to form. Additionally, the simulations were less reliable for higher doses, where the large number of replacement collision events at high vacancy concentration were not taken into account. One possibility for an improvement was to measure the implanted atom density by secondary ion mass spectrometry after the FIB \st{irradiation}\hl{implantation} and before the \SI{900}{\celsius} annealing~\cite{dhaenens-johanssonOpticalPropertiesNeutral2011}. Near-surface NV centers also had different collection efficiencies form the deep NVs that are used for PL normalization~\cite{santoriVerticalDistributionNitrogenvacancy2009}. Our calculation did not take this into consideration so might result in a slightly larger formation yield.

\section{Inhomogeneous Broadening}\label{cryogenicPL}

\begin{figure}[ht!]
    \centering
    \includegraphics[scale=1]{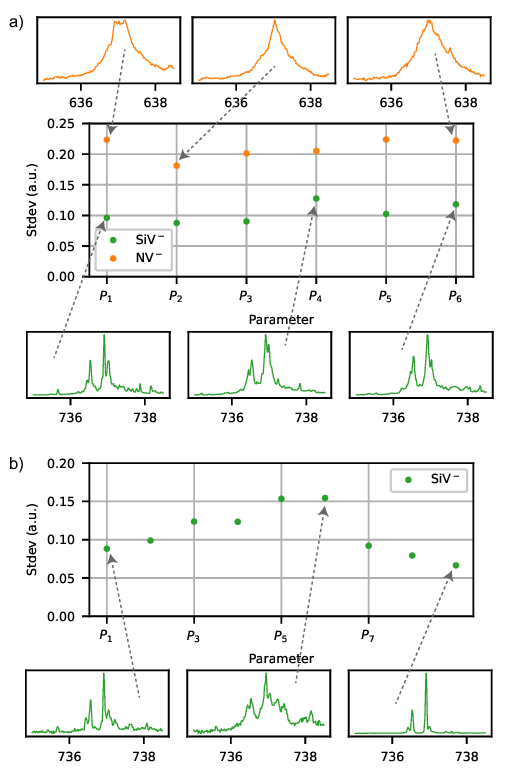}
    \caption{Inhomogeneous broadening and example PL, For each defect, the spectra from both the highest-dose and the lowest-dose squares were displayed; spectra with both the largest standard deviation and the smallest standard deviation were displayed. Plot (a) was measured from the SiN$_x$ sample; plot (b) was measured from the SiO$_2$ sample} 
    \label{inhomogeneous}
\end{figure}

As discussed in the main article, low-temperature (10\,K) PL spectra were collected from every \st{irradiation}\hl{implantation} square to measure the inhomogeneous broadening of both NV$^-$ and SiV$^-$ centers. To compare these spectra, we normalized each spectrum with their intensity and plotted the standard deviation in FIG.~\ref{inhomogeneous}. There was no clear dependence of the inhomogeneous broadening on the Ga$^+$ ion fluence. The representative spectra shown in FIG.~4 in the main article were the ones with the smallest standard deviation in the plots.

\section{Charge Stability}\label{charge}

\begin{figure}[hbtp!]
    \centering
    \includegraphics[scale=1.2]{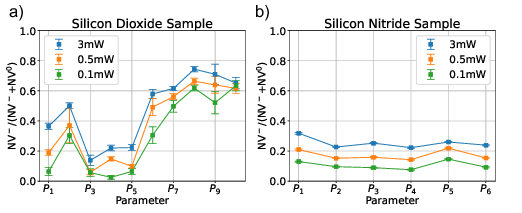}
    \caption{Charge state ratio of NV center(s) inside \st{irradiation}\hl{implantation} squares from both samples. Each data point was calculated by measuring the NV emission spectrum at 10\,K and fitting the NV$^-$ and NV$^0$ ZPL on a quadratic background. Error bars in the plots were from fitting uncertainties. P$_1$ to P$_{10}$ are the FIB parameters as described in TABLE~\ref{doseTab}.} 
    \label{chargePlot}
\end{figure}

One way of probing the irradiation-induced damage and the effects of the introduced impurities was by looking at the ratio between NV$^-$ and NV$^0$. Ideally, the proportion of NV$^-$/(NV$^-$+NV$^0$) should reach at least 70\%~\cite{aslamPhotoinducedIonizationDynamics2013,santoriVerticalDistributionNitrogenvacancy2009,fuConversionNeutralNitrogenvacancy2010} and NV$^-$ should remain dominant in the limit of low optical excitation power. To measure this ratio, we collected the NV spectrum at 10\,K from every \st{irradiation}\hl{implantation} square on both samples with three different 532\,nm excitation powers. We then spectrally separated the NV$^-$ and NV$^0$ ZPL from their background using the method described in Section \ref{fitting} and calculated their intensity ratio. As shown in FIG.~\ref{chargePlot}(b), the ratio was less than 35\% for all SiN$_x$ squares and decreased with the excitation power. While the power dependence implied a strong interaction between the NV centers and the near-by electron donors~\cite{gaebelPhotochromismSingleNitrogenvacancy2006}, two factors were expected to have caused this low ratio. First, the irradiation-induced damage which created sp$^2$ carbons and electron traps at the diamond surface would decrease the ratio by limiting the available charge carriers around the NV centers, especially when these NV centers were very close to the sample surface~\cite{rondinSurfaceinducedChargeState2010}. Second, the implanted electron acceptors, such as the substitutional gallium defects, would compensate for the available donors (mostly nitrogen atoms in this case) around the NV centers~\cite{santoriVerticalDistributionNitrogenvacancy2009,gossVacancyimpurityComplexesLimitations2005a}. This was partially confirmed by the lower-dose squares (P$_8$,P$_9$,P$_{10}$) on the SiO$_2$ sample (FIG.~\ref{chargePlot}(b)), where a larger ratio with weaker power dependence was observed. In these squares, since lower gallium ion fluence was applied by reducing the FIB dwell time, both the surface damage and the number of implanted acceptors were reduced. Although we did not expect any NV center formation by recoil implantation in the SiO$_2$ sample, we managed to find at least one location inside each square where we could detect the NV PL from NV center(s) that was formed by the naturally abundant nitrogen impurities in the sample~\cite{vandamOpticalCoherenceDiamond2019a}. For this reason, the variance shown by the data from the SiO$_2$ sample was larger than that from the SiN$_x$ sample, as each NV center had a different micro-environment. Nevertheless, these data were still good qualitative representations of the environment inside each \st{irradiation}\hl{implantation} square.

\section{Spin Coherence}\label{odmr}
\begin{figure}[hbpt]
    \centering
    \includegraphics[scale=1]{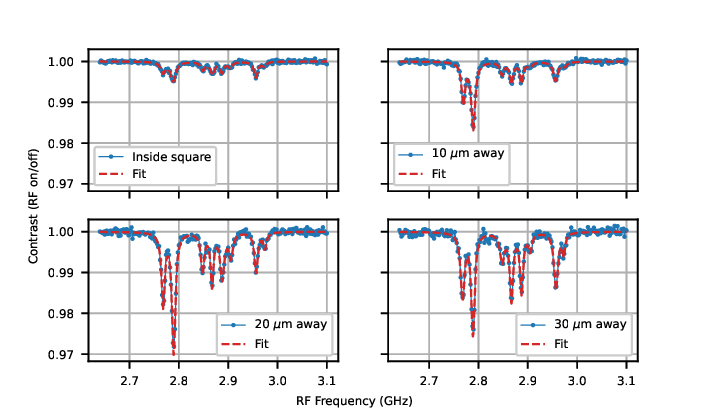}
    \caption{ODMR spectra from the four locations around the highest-dose \st{irradiation}\hl{implantation} square on the SiN$_x$ sample.} 
    \label{odmrplot}
\end{figure}

The continuous wave optically detected magnetic resonance (CW-ODMR) experiments in this work were performed by sweeping a radio-frequency (RF) wave around the NV zero-field splitting frequency (2.87\,GHz), while monitoring the NV PL intensity change. The RF signal was delivered to the sample by the same microwave antenna as described in Ref.~\cite{sasakiBroadbandLargeareaMicrowave2016}, where measurements and simulations of the S$_{11}$ curve for this
antenna design were provided. In particular, there was a $\sim$\,4dB increase of the S$_{11}$ curve from the resonance frequency (2.79\,GHz) to 3.0\,GHz. This explained the asymmetric spectrum intensity as shown in FIG.~\ref{odmrplot}, where the high-frequency resonance-dips displayed smaller contrast than the lower-frequency ones. 

\begin{figure}[htpb]
    \centering
    \includegraphics[scale=1]{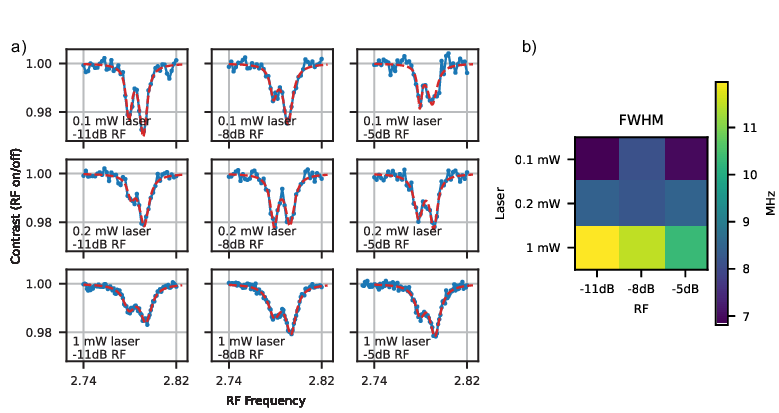}
    \caption{(a) CW-ODMR spectra taken at $\SI{20}{\micro\metre}$ away from the square center using different RF and excitation powers. The unit for the x-axis in GHz. Only relative unit (dB) for RF power was used in this work because of the complication in calculating the actual power delivered to the sample. The RF power for spectra shown in FIG.~\ref{odmrplot} was -10\,dB. (b) Average FHWM of the two peaks from each of the 9 data shown in (a).} 
    \label{power}
\end{figure}

For the four ODMR spectra shown, the excitation beam was inside the square, $\SI{10}{\micro\metre}$, $\SI{20}{\micro\metre}$, and $\SI{30}{\micro\metre}$ away from the center of the square, respectively. Each spectrum was fitted with eight individual Lorentzian functions to determine an average linewidth (FWHM) of 10.3$\pm$1.2 MHz, 9.5$\pm$1.4 MHz, 8.37$\pm$0.88 MHz, and 8.5$\pm$1.3 MHz (from center to $\SI{30}{\micro\metre}$ away). The uncertainties in the linewidths were dominated by the variation of FWHMs among different resonance dips, which came from the antenna response at different frequencies. We also calculated four zero-field spitting frequencies for each spectrum from the four pairs of resonances, and determined the mean and the standard deviation to be:  2.8741$\pm$0.0029\,GHz, 2.8744$\pm$0.0020\,GHz, 2.8746$\pm$0.0026\,GHz, and 2.8742$\pm$0.0027\,GHz (from square-center to $\SI{30}{\micro\metre}$ away). 

A relatively high laser power (2\,mW) was used for these measurements to achieve a higher NV$^-$/NV$^0$ ratio (as implied by FIG.~\ref{chargePlot}(b)). However, this power also broadened the linewidth of the ODMR spectra. As shown in FIG.~\ref{power}, the linewidth decreased as the laser power was reduced. The narrowest linewidth (6.81\,MHz) was measured with 0.1\,mW excitation and -11\,dB RF power. The fit of this spectrum showed a contrast of 2.9\% for the deepest resonance dip. This contrast was mostly limited by the NV$^-$/NV$^0$ ratio (FIG.~\ref{chargePlot}) and the other non-NV contributions to the total PL intensity (FIG.~\ref{ratio}). Specifically, we have measured the ratio of NV PL contribution to the total PL intensity to be 0.81, and that the ratio NV$^-$/(NV$^-$+NV$^0$) was below 0.5 at 0.1\,mW. Therefore, the optimal ODMR contrast from these NV centers can be calculated by considering the ideal case where all fluorescence comes from the NV emission, and the NV$^-$/(NV$^-$+NV$^0$) ratio was larger than 0.7. In this case, the contrast depth of the deepest resonance dip would be at least $2.9\%*0.7/(0.5*0.81) = 5\%$. Hence, considering the four different crystallographic orientations of NV centers in diamond, we could estimate the total contrast to be at least 20\%, which was consistent with the result for CW-ODMR in Ref.~\cite{dreauAvoidingPowerBroadening2011a}. This confirmed that the limiting factors for the ODMR PL contrast were indeed what we expected in the main article.

\section{\hl{Defect Localization}}\label{simulation}
\begin{figure}[htpb]
    \centering
    \includegraphics[scale=1.8]{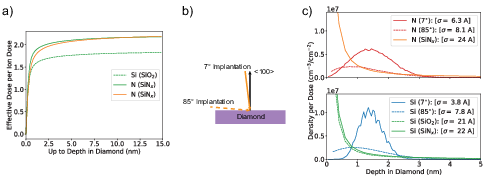}
    \caption{\hl{Depth distribution of defects introduced by recoil implantation and direct ion implantation. (a) Cumulative density of the recoil-implanted atoms by simulations that were shown in FIG.~3 of the main article. The x-axis represents the vertical distance considered from the diamond surface up to a certain depth for the cumulative density. (b) Illustration diagram of different angle ion implantation. (c) TRIM simulation of the impurity distributions implanted by different methods but with the same average depth of 1.4\,nm. The straggle for each implantation are labeled in the legend of the plot. The straggle here is defined as the standard deviation of the distribution.}} 
    \label{straggling}
\end{figure}

\hl{
Because FIG.3(a) and (b) in the main article are not intuitive for estimating the density of the recoil implanted silicon and nitrogen atoms, we integrated over the density-per-dose profile of the atom depth distribution to obtain a cumulative atom-density per ion-density function of depth. This function, as plotted in FIG.~\ref{straggling}(a), allows for better  estimating the density of the atoms within a certain depth in diamond. For example, in the top 4 nm of the diamond, the density of silicon and nitrogen atoms in the SiN$_x$ sample is approximately twice the gallium ion dose, while the silicon atoms in the SiO$_x$ sample is approximately 1.6 times the gallium ion dose.

To compare recoil implantation to direct ion implantation, we performed SRIM simulations for nitrogen and silicon ion implantation with both normal (7 degree) and large incident angle (85 degree) into diamond. As shown by the results in FIG.~\ref{straggling}(c), directly implanted ions has smaller straggling ($<1$\,nm) than the recoil-implanted atoms ($\approx2$\,nm).}

%We also calculated the average number of vacancies per recoil-implanted atom (64 for SiO$_2$ and 70 for SiN$_x$) and compared it to the number of vacancies per directly implanted ion (140 for 30\,keV nitrogen ion and 249 for 30\,keV silicon ion).

\bibliography{sm_Book}